# Characterization of a gigabit transceiver for the ATLAS inner tracker pixel detector readout upgrade

**C. Chen,**[a,b] **D. Gong,**[b,*] **D. Guo,**[a] **G. Huang,**[a] **X.Huang,**[a,b] **S. Kulis,**[c] **P. Leroux,**[d]
**C. Liu,**[b] **T. Liu,**[b] **P. Moreira,**[c] **J. Prinzie,**[d] **Q. Sun,**[b] **P. Wang,**[b] **L. Xiao,**[a] and **J. Ye**[b]

[a] *Central China Normal University,*
  *Wuhan, Hubei 430079, P.R. China*
[b] *Southern Methodist University,*
  *Dallas, TX 75275, USA*
[c] *CERN,*
  *1211 Geneva 23, Switzerland*
[d] *KU Leuven,*
  *Leuven, Belgium*

  *E-mail:* dgong@mail.smu.edu

ABSTRACT: We present a gigabit transceiver prototype Application Specific Integrated Circuit (ASIC), GBCR, for the ATLAS Inner Tracker (ITk) Pixel detector readout upgrade. GBCR is designed in a 65-nm CMOS technology and consists of four upstream receiver channels, a downstream transmitter channel, and an Inter-Integrated Circuit (I2C) slave. The upstream channels receive the data at 5.12 Gbps passing through 5-meter 34-American Wire Gauge (AWG) Twin-axial (Twinax) cables, equalize them, retime them with a recovered clock, and then drive an optical transmitter. The downstream channel receives the data at 2.56 Gbps from an optical receiver and drives the cable as same as the upstream channels. The jitter of the upstream channel output is measured to be 35 ps (peak-peak) when the Clock-Data Recovery (CDR) module is turned on and the jitter of the downstream channel output after the cable is 138 ps (peak-peak). The power consumption of each upstream channel is 72 mW when the CDR module is turned on and the downstream channel consumes 27 mW. GBCR survives the total ionizing dose of 200 kGy.

# Contents

# 1. Introduction

In the ATLAS inner tracker (ITk) Pixel Detector Phase-II upgrade [1]，it is a challenge to design the readout system because the electronic modules must be low mass, radiation hard, halogen free, and easy to operate. Because there is no common ground reference among the serial-powered front-end (FE) modules [2], each of which is modeled as a 150-μm-thick silicon wafer, a 1-μm-thick copper layer, and one 20-μm-diameter Sn-Ag bump-bond per pixel channel, it is preferable to choose optical data transmission to get rid of potential grounding issues. Because it is impossible to place the laser diodes into a sufficient radiation environment, in the ITk readout system, an optical box, including optical transceiver modules VTRx+ [3], is mounted at Patch Panel 1 (PP1), which is about 5 meters away from the detector. As shown in figure 1, each FE readout module has four serial output links, each operating at 1.28 Gbps. The Application Specific Integrated Circuit (ASIC) RD53B [4] transmits these signals out of the detector. An ASIC called Aggregator aggregates these four links into a single fast data line. The fast data line directly passes through a 5-meter 34-American Wire Gauge (AWG) twin-axial (Twinax) cable. In order to compensate the high-frequency loss induced by the long cable, an ASIC called GBCR is being developed. GBCR recovers the data and transmits them to the VTRx+ modules in the optical box. The VTRx+ modules send the data to the counting room through 80-meter fibers. Both Aggregator and GBCR are mounted on pluggable substrates. The two ASICs and the cable together are called an active cable. In this paper, we present the design and test results of GBCR.

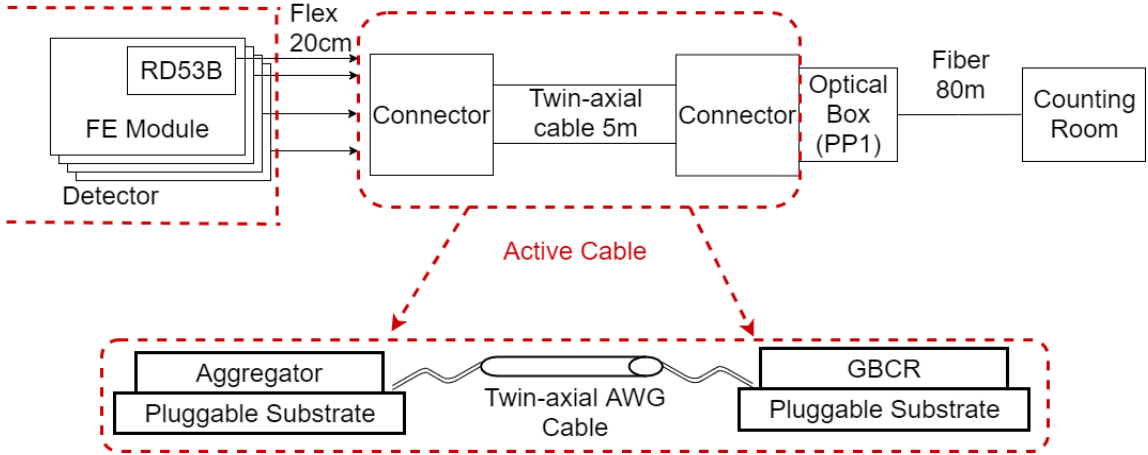

Figure 1. Block diagram of the ITk readout chain.

## 2. Design

### 2.1 Overall structure

Figure 2 is the block diagram of GBCR. This chip includes four upstream channels and one downstream channel. Among upstream channels, there are three baseline channels and one test channel. Each baseline channel consists of an input stage (equalizers and Limiting Amplifiers (LAs)), a Clock-Data-Recovery (CDR) module, and a Current-Mode-Logic (CML) driver. The test channel has an additional test structure, which is beyond the scope of this paper. All upstream channels share an Automatic-Frequency-Calibration (AFC) module [5] in order to calibrate the Voltage Control Oscillator (VCO) [6] in each channel. The upstream channel takes the 5.12-Gbps data from the Aggregator through a 5-m Twinax cable as input signals and sends the recovered data to the VTRx+ module. The downstream channel consists of an LA, a passive attenuator, a pre-emphasizer, and a CML driver. The downstream channel receives the 2.56-Gbps data from the VTRx+ module and transmits the data to the Aggregator through the same cable as the upstream channels. An I$^2$C slave provides the control signals of all channels.

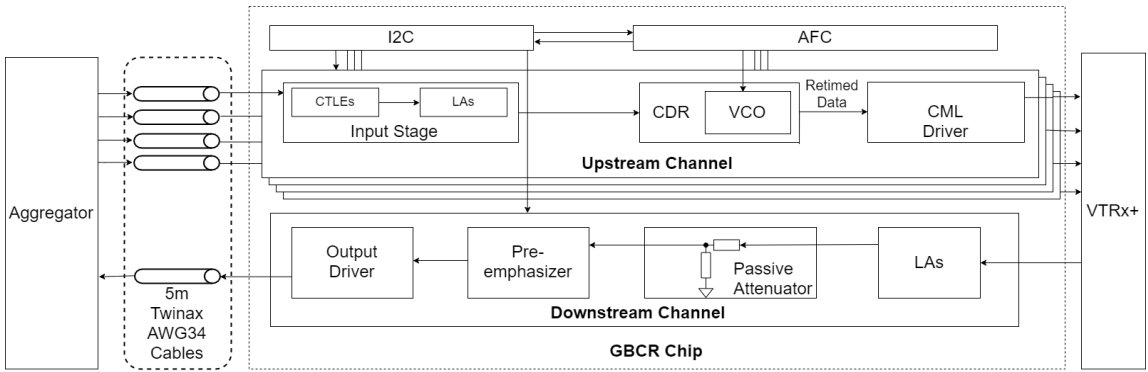

Figure 2. Block diagram of GBCR.

The input signals of the upstream channels of GBCR come from the aggregator with an output amplitude of larger than 200 mV (peak-peak). The signals are encoded in the 64b/66b protocol with a total jitter of less than 100 ps (peak-peak). The output amplitude of the upstream channels of GBCR must be larger than 400 mV (peak-peak). GBCR must tolerate a Total ionizing dose (TID) of 200 kGy.

## 2.2 Upstream Channels

### 2.2.1 Input stage

When a high-speed signal passes through a long cable, its high-frequency components attenuate more than its low-frequency components, resulting in the signal quality deterioration. An equalizer improves the signal quality because it behaves like a high-pass filter and degenerates the low-frequency components [7]. The input stage has two stages of Continuous-Time Linear Equalizers (CLTEs). Figure 3(a) is the schematic of a CTLE. Each CTLE has a zero and a pole as follows:

$$\omega_Z = \frac{1}{R_S C_S} \tag{1},$$

$$\omega_{p1} = \frac{1 + \frac{g_m R_S}{2}}{R_S C_S} \tag{2}.$$

The equalization strength is described as

$$Equalization\ Strength = 1 + \frac{g_m R_S}{2} \tag{3}.$$

We adjust the equalization strength by tuning Cs and Rs, based on the high-frequency loss of the long cable and the data rate.

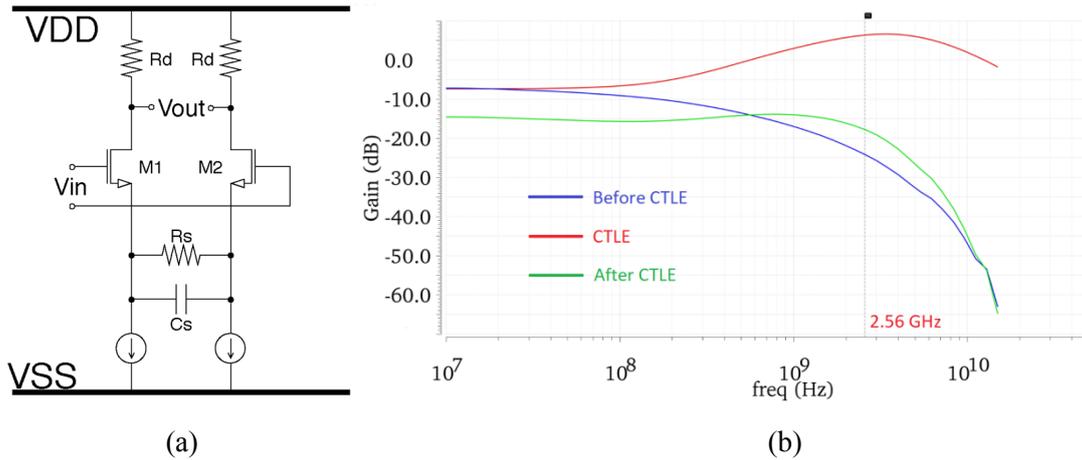

(a)                                        (b)

Figure 3. Schematic of a CTLE (a) and AC response of the equalizer (b).

Figure 3(b) is the overall AC response of the equalizers. The blue curve is the measured frequency response of the Twinax cable. The frequency component at 2.56 GHz is attenuated 16.5 dB more than at DC. The red curve is the AC response of the equalizers. The equalization strength is 15.0 dB. The green curve is the AC response of the signal out of the equalizers. As can be seen in the figure, the green curve is flatter than the blue curve. The bandwidth of the equalizers is about 3.2 GHz.

The input stage includes three stages of LAs [8] to amplify the signal amplitude to meet the input requirements of the next stage. The schematic of each stage is shown in Figure 4(a). The gain of each stage is described as

$$Gain = g_m R_d \quad (4).$$

The simulation results of the overall three stages of LAs are shown in Figure 4(b). The three stages of LAs provide a gain of 25 dB. The overall bandwidth of LAs is 3.3 GHz.

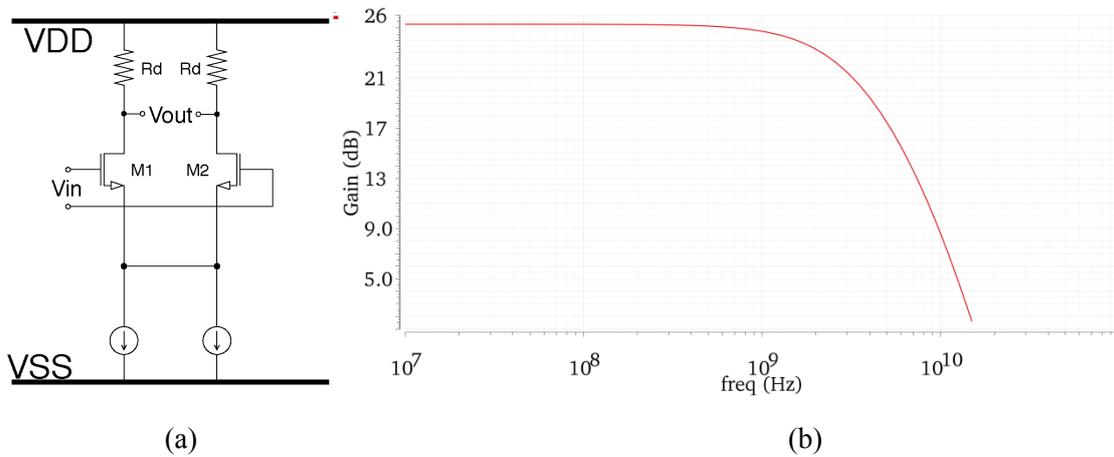

(a)                                    (b)

Figure 4. Schematic of an LA stage (a) and overall AC response of three stages (b).

### 2.2.2 CDR

Figure 5 is the block diagram of the CDR module and the AFC module. The CDR module and the AFC module are migrated from lpGBT [9]. The basic structure of the CDR is shown in the dotted-line frame. The CDR includes a phase detector (PD), a charge pump (CP), a low-pass filter (LPF), and an LC-tank VCO [10]. The PD detects the phase difference between the 5.12-GHz clock from the VCO and the 5.12-Gbps data from the equalizers. The CP converts the phase difference into a current signal. The LPF filters the high-frequency components of the current signal and integrates the remaining low-frequency components into the control voltage of the VCO. The VCO adjusts its output frequency to follow the frequency of the input data until the two phases are close. The CDR samples the data with the VCO output clock [11].

The CDR module implements an LC-tank VCO. Since the tuning range of such a VCO is only about 10-15% of the central frequency, a series of capacitors parallel to the varactor are implemented in the VCO. The frequency of the VCO can be automatically adjusted by switching different capacitors [12]. The frequency adjustment process is implemented via the AFC module. Figure 5 shows the data flow between the AFC module and the CDR module. Each upstream channel has two dividers, a data divider and a clock divider. The divided-by-32 data divider obtains a clock close to 40 MHz from the 5.12-Gbps data, while the divided-by-128 clock divider acquires another 40-MHz clock from the VCO output clock, which is about 5.12

GHz. The AFC module ensures the output frequency of the VCO to be close to 5.12 GHz by one of these two 40 MHz clocks.

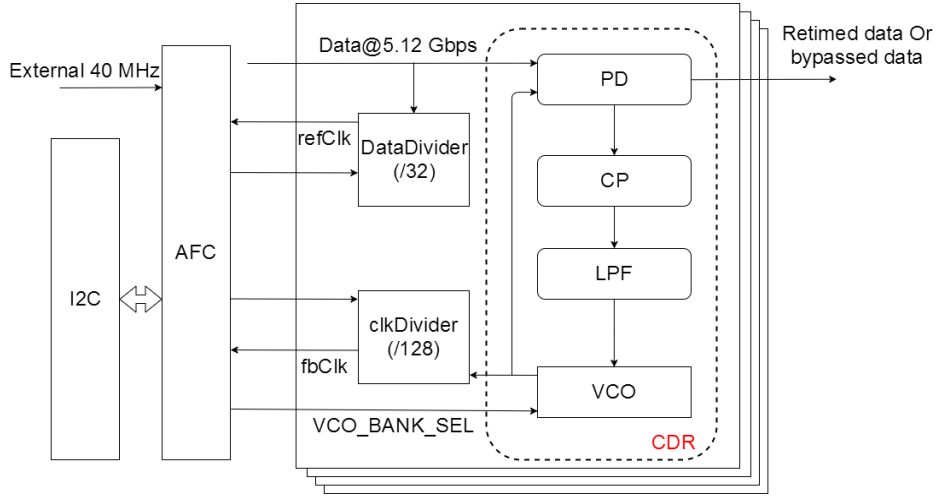

Figure 5. Block diagram of the CDR and AFC modules.

### 2.2.3 CML driver

Figure 6 is the schematic and the AC response of the CML driver [14]. The output driver has the same structure as an LA stage except that the CML driver has two 50-Ω pull-up resistors. The gain of the output stage is -0.46 dB with the bandwidth is 6.1 GHz at the typical process corner, 27°C, and the nominal power voltage.

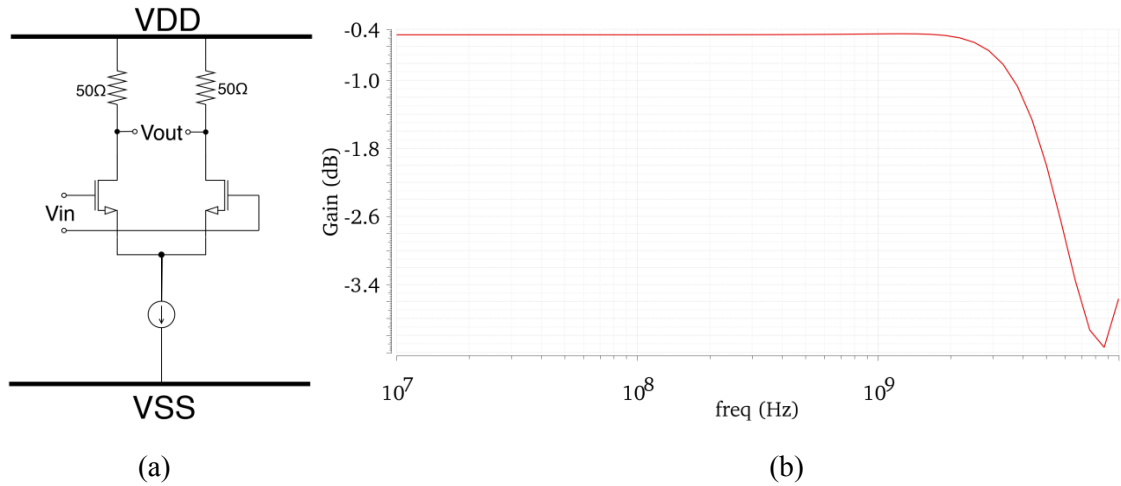

(a)                                                        (b)

Figure 6.  Schematic of the driving stage (a) and AC response of the whole CML driver (b).

### 2.3 Downstream Channel

The downstream channel takes the 2.56 Gbps data from the VTRx+ as the input source and then pre-emphasizes [15] them to meet the requirements of the Aggregator. The amplitude of these differential signals is from 75 mV to 1V (peak-peak). As shown in Figure 2, this channel consists of several stages of LA, a passive attenuator, a pre-emphasizer, and an output driver. Among these modules, the 3 stages of LA provide 18.4 dB gain to amplify the input signal to saturation even when the signal amplitude is the minimum to avoid the eye distortion. The gain of the passive attenuator is -3.5 dB to attenuate the maximum amplitude signals for the next stage. The pre-emphasizer, which is adopted the CTLE structure, is used to pre-emphasize the signal to compensate for the high-frequency loss due to long-distance transmission in advance. The last output driver stage drives the long cable and sends the handled data to the Aggregator as well. Both the LAs and CTLE have the same structure as discussed in the sections above. Because of the low data rate requirements, the parameters in these circuits are different from the ones used in the upstream channel.

## 2.4 I$^2$C module and overall layout

An I$^2$C slave module with 256 binary bits is included in GBCR. The I$^2$C slave module is implemented in Triple Modular Redundancy (TMR) to avoid the effects of Single Event Upset (SEU).

The total size of the GBCR die is 2 mm by 3 mm. The layout of GBCR is shown in Figure 7.

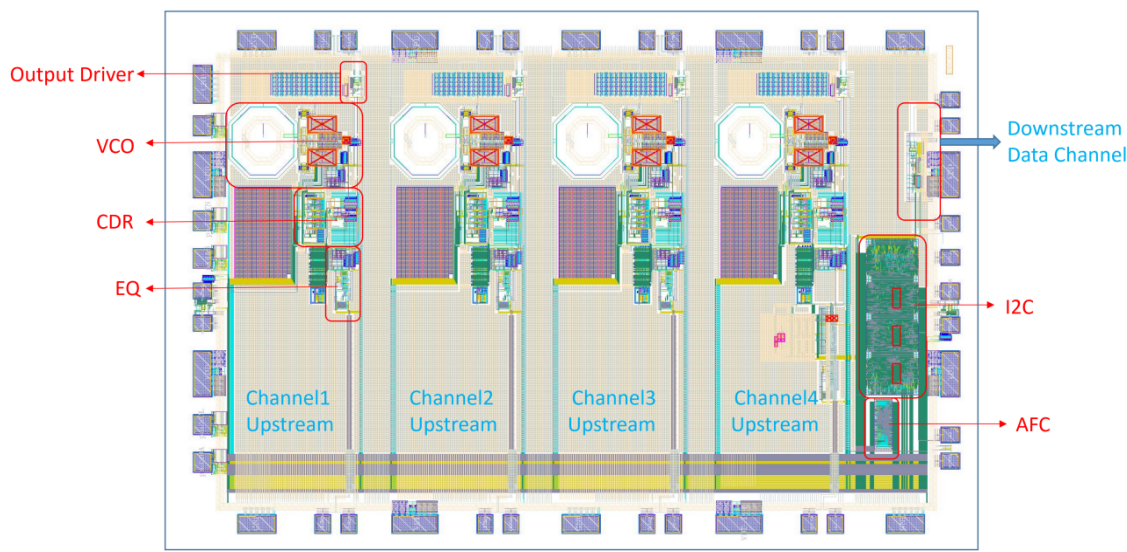

Figure 7. Screenshot of overall layout.

## 3. Test results

### 3.1 Testbench

Figure 8(a) shows the block diagram of the test bench. For the upstream channels, a signal generator (Model: CENTELLAX TG1C1-A with a clock module CENTELLAX PCB12500) provides 5.12-Gbps pseudorandom-binary-sequence (PRBS) $2^{31}$-1 data with an amplitude of 200 mV (peak-peak) passing through a 6-m Twinax cable as the input of a GBCR test board and the trigger signal for the oscilloscope (Model: Tektronix DSA70804B). The GBCR test board is directly connected to the oscilloscope by two 15 cm coaxial cables through a differential SMA probe (Model: Tektronix P7313SMA). For the downstream channel, the Twinax cable is after the GBCR test board. Figure 8(b) is a picture of the test bench of the downstream. Two packaged GBCR chips are shown in figure 8(b).

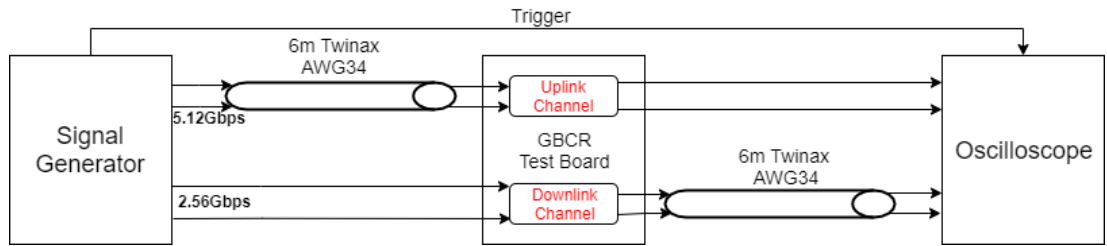

(a)

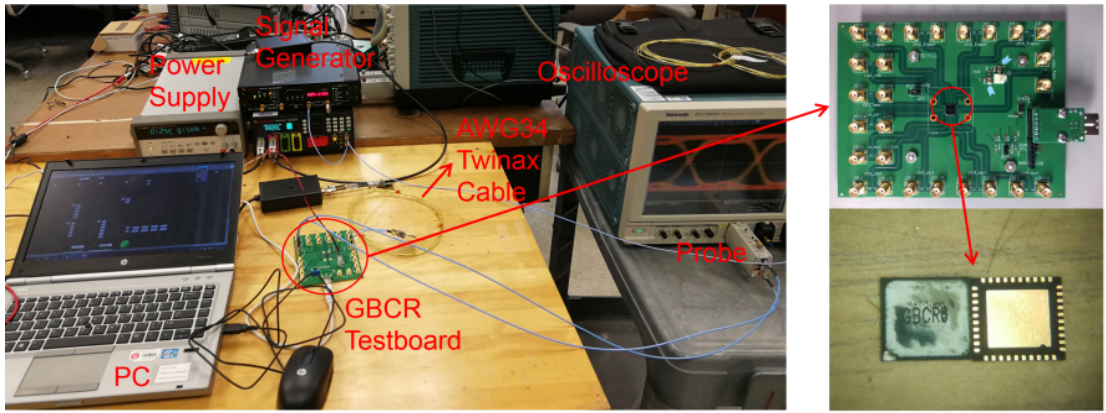

(b)

Figure 8. Block diagram (a) and picture of testbench (b).

### 3.2 Upstream channels

Figure 9 shows the eye diagrams at different nodes of an upstream channel. The eye diagram of the output of the the signal generator is shown in figure 9(a). After the signal passes through the 6-m Twinax cable, the eye diagram is fully closed, as shown in figure 9(b). After recovered by the equalizers and the CDR, the eye diagrams are open, as shown in figures 9(c) and 9(d). When the CDR is turned off, the total jitter is 79 ps (peak-peak) and the random jitter is 4.3 ps (root mean square (RMS)) under the resistance setting of 0 and the capacitance setting of 3 of the equalizers. When the CDR is turned on, the total jitter decreases to 35 ps (peak-peak) and the random jitter is 1.5 ps (RMS).

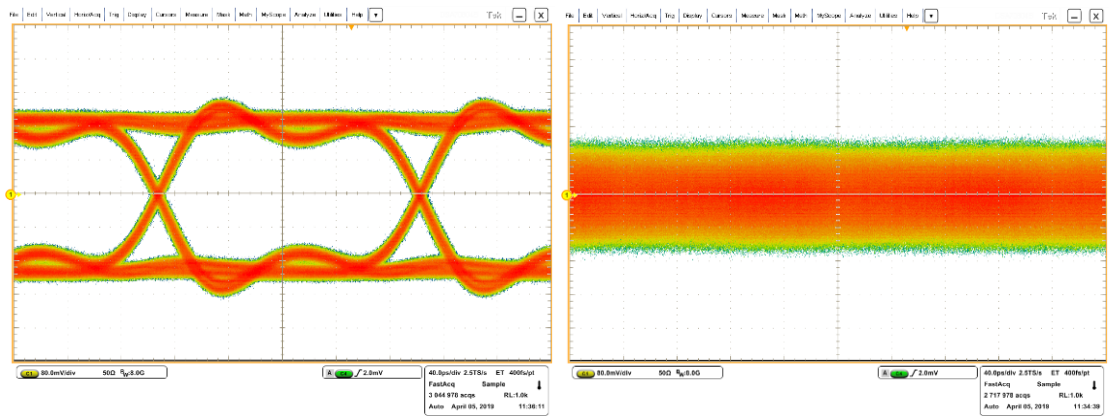

(a)                                                     (b)

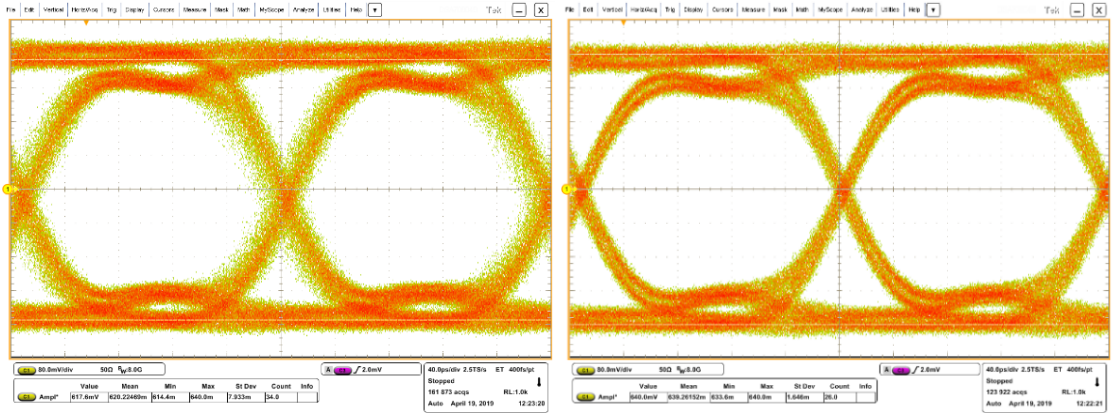

(c)                                                    (d)

Figure 9. Eye diagrams after the signal generator (a), after the 6m Twinax cable (b), after the whole upstream when CDR is disabled (c), and after the whole upstream when CDR is enabled (d).

As mentioned in Section 2.2.2, the equalization strength can be adjusted by tuning resistance and capacitance in the CTLEs. We measured the jitter in different equalization settings. The results are shown in Figure 10. We didn't distinguish the random jitter and the deterministic jitter because of limited test time. Instead, we measured the histogram of the zero-crossing time and estimated the jitter. The jitter is close to 5 ps (RMS) when the resistance setting is 0 and the capacitance setting is 3.

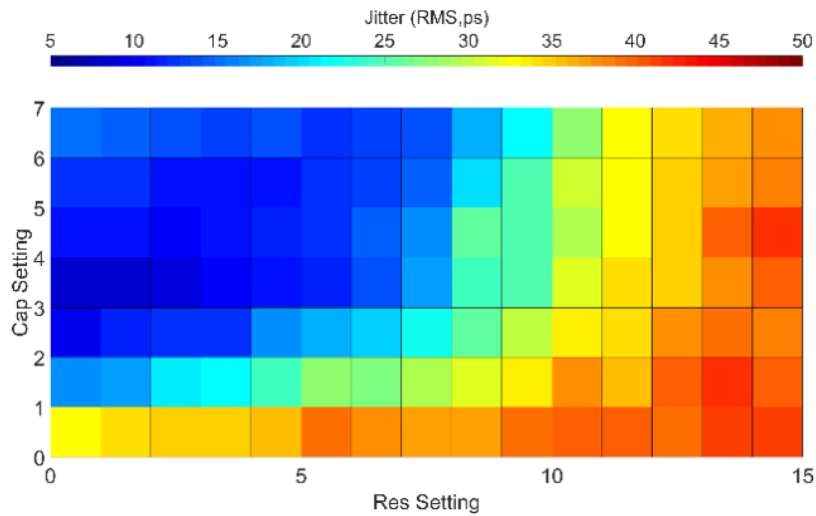

Figure 10. Relationship between equalization and jitter.

Figure 11 is the bit error rate (BER) at various resistance settings in CTLEs (the capacitance setting is fixed to 4). When the resistance setting in CTLEs was less than or equal to 7, no bit error occurred among the test duration of 20 minutes. As the resistance setting grew, the BER increased and finally saturated at 0.5. These results indicate that as far as the equalizers are correctly configured, the upstream recovers the data correctly.

Figure 12 is the results of the input sensitivity test. As shown in the figure, when the input amplitude was larger than 200 mV, the total jitter was stable at about 80 ps (peak-peak), but the output amplitude was stable at 614 mV, independent on the input amplitude. The results indicate that we achieve an input sensitivity of 200 mV.

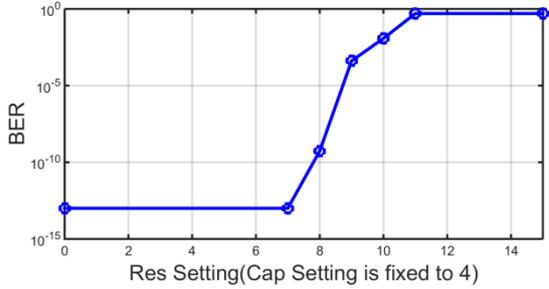

Figure 11. BER Test.

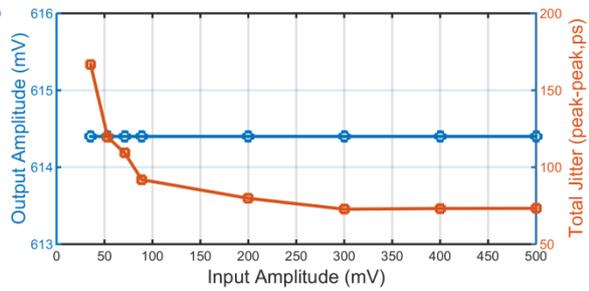

Figure 12. Sensitivity Test.

The curve shown in figure 13(a) is the relationship between the VCO output frequency and the capacitance setting of the VCO. When we took the internal clock as the calibration source, after running the AFC 1000 times, we obtained the histogram of the calibrated capacitance setting shown in figure 13(b). The calibrated capacitance setting could be 13, 14, 15, or 16 because the tiny difference may exist between the recovered clock and the data source. All the output frequencies corresponding to these 4 codes are close to 5.12 GHz. The most possible capacitance setting is 15. If the calibration clock and the original data are synchronous, the difference mentioned above does not exist anymore. In this case, when we used the external clock generated from the original data source as the calibration source, we obtained the only code 15. These results indicate that that the AFC module works well as expected.

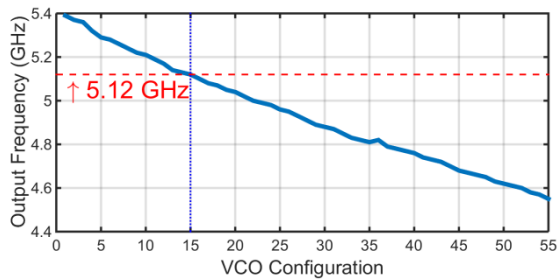

(a)

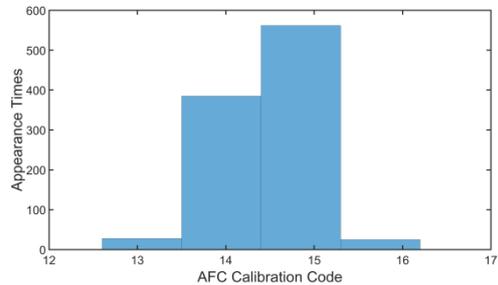

(b)

Figure 13. Relationship between the VCO capacitors setting and the VCO output frequency (a) and the histogram of calibration code distribution (b).

The power consumption of each upstream channel is measured to be about 29 mW when the CDR is turned off and 72 mW when the CDR is turned on, respectively. The measured power consumptions match the simulation values.

### 3.3 Downstream channel

Figure 14 shows the eye diagrams of the downstream channel before and after the Twinax cable. The overshoot, as shown in figure 14(a), in the eye diagram before the 6-m Twinax is obvious because of pre-emphasis. After the 6-m Twinax cable, the eye diagram, as shown in figure 14(b), is still open. The total jitter is 138 ps (peak-peak) and the random jitter is 4.1 ps (RMS). Because the length of test cable and some layout negligence, the total jitter is larger than we expected. The power consumption of the downstream channel is about 27 mW. The measured power consumption matches the simulation value.

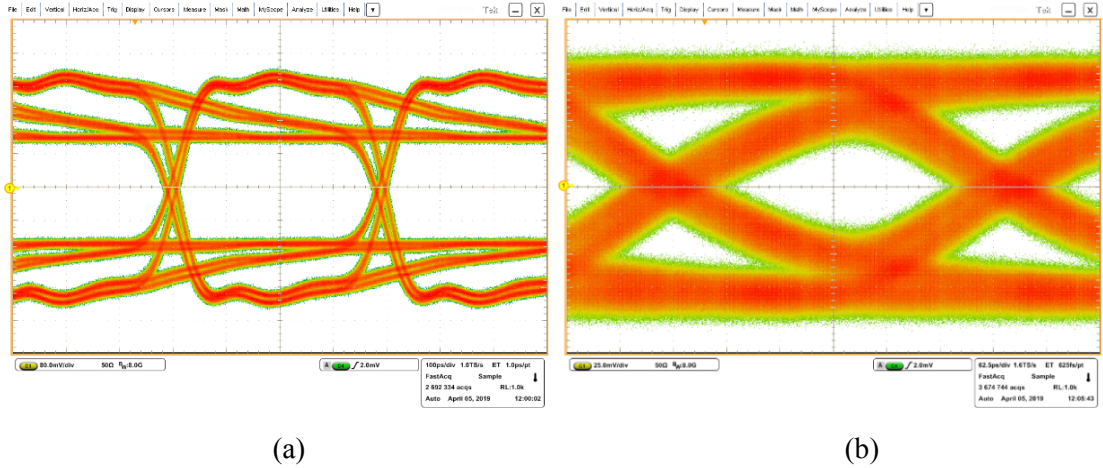

(a)                                                          (b)

Figure 14. Eye diagrams before (a) and after the Twinax cable (b).

### 3.4 Irradiation tests

One GBCR chip was irradiated in Cobalt-60 gammy rays for 31 days and accumulated a total ionizing dose of 200 kGy. During irradiation, the chip was powered up. After irradiation, the chip still functioned properly and no significant performance degradation was observed. A single event effect test will be conducted in the future.

## 4. Summary and outlook

We have designed and tested a gigabit transceiver prototype ASIC called GBCR fabricated in a 65-nm CMOS technology for the ATLAS ITk detector readout upgrade. For the upstream channels, when the CDR is turned on, the output total jitter is about 35 ps (peak-peak). For the downstream channel, the total jitter after the cable is 138 ps (peak-peak). Each upstream channel consumes 72 mW when the CDR is turned on and the downstream channel consumes 27 mW. GBCR survives 200 kGy in the irradiation test.

The second prototype called GBCR2 has been submitted. In this new design plan, the ASIC Aggregator has been removed, and GBCR2 is straightly connected to the RD53B in the detector through a combo that combines a flex cable and a Twinax cable. Compared with GBCR1, the data rate decreases to 1.28 Gbps and the number of upstream channels increases to 7 to match the group number of e-links in lpGBT. Two downstream channels transmit the 160-Mbps data from lpGBT to RD53B. GBCR2 will be tested in 2020.

## Acknowledgments

This project was supported by the US-ATLAS Phase-2 upgrade grant administrated by the US-ATLAS Phase-2 Upgrade Project Office and the Office of High Energy Physics of the U.S. Department of Energy under contract DE-AC02-05CH11231. Thanks so much to Drs. Andrew Young and Dong Su for providing the cable model. We appreciate to Dr. Maurice Garcia-Sciveres and Ms. Veronica Wallangen with Lawrence-Berkeley National Laboratory (LBNL) for lending us various 34AWG cables and interface boards. The authors also would like to thank Mr. James Kierstead from Brookhaven National Laboratory (BNL) for the irradiation test. It is highly helpful to us as many institutes have made great contributions to the design and test of this chip.